# Po-production in lead: A benchmark between Geant4, FLUKA and MCNPX

Alfredo Ferrari[1], Daniela Kiselev[2], Tatsumi Koi[3], Michael Wohlmuther[2] and Jean-Christophe Davide[4]
[1]CERN, European Laboratory for Particle Physics, Switzerland
[2]Paul Scherrer Institut, Switzerland
[3]SLAC National Accelerator Laboratory, USA
[4]CEA-Saclay, France

**Abstract**

*On the last SATIF a comparison between the measured activities of the polonium isotopes Po-208, Po-209 and Po-210 and the simulated results using the particle transport Monte Carlo code MCNPX2.7.0 was presented. The lead samples were cut from the SINQ spallation target at the Paul Scherrer Institut (PSI) and irradiated in 2000/2001 by 575 MeV protons. The Po-isotopes were separated from the samples using radiochemical methods by the group of Dorothea Schumann at PSI and measured with an alpha sensitive detector. Choosing the default model in MCNPX, Bertini-Dresner, the prediction underestimated the measured activities by up to several orders of magnitude. Therefore the Liège intranuclear-cascade model (named INCL4.6) coupled to the de-excitation model ABLA07 were implemented into MCNPX2.7.0 and very good agreement was found to the measurement. The reason for the disagreement was traced back on one hand side to a suppression of alpha reactions on the lead isotopes leading to Po and on the other hand side neglecting the triton capture on Pb-208, which leads to Pb-210 and decays into Po-210 with a much longer life time (22.3. years) than the decay of Po-210 itself (138 days).Therefore the activity of Po-210 was 10 years after the irradiation much higher than without the reaction channel via Pb-210.*

*The prediction of the Po-isotope activities turns out to be a sensitive test for models and codes as it requires the accurate treatment of reaction channels not only with neutrons, protons and pions but also with alphas and tritons, which are not considered in intra-nuclear cascade models of the first generation. Therefore it was decided to perform a benchmark by comparing the results obtained with MCNPX2.7.0 using INCL4.6/ABLA07 to the predictions of FLUKA and Geant4. Since the model of the SINQ spallation source requires an elaborate geometry and due to the small size of the samples statistical relevant results are very difficult to obtain, a toy model was setup. The toy model has a simplified geometry preserving the main features of the original geometry. The results for the activities of the three Po-isotopes and Pb-210 as well as the energy spectra for alphas and tritons obtained with the three particle transport Monte Carlo codes will be presented.*

**Motivation and Introduction**

When radioactive material has to be disposed, the authorities in Switzerland require a complete nuclide inventory for long-lived isotopes ($T_{1/2} > 60$ d). These predictions have to be validated as experimental data become available. The authorities are particularly interested in $\alpha$-emitters as these isotopes are more difficult to detect and therefore classified as more risky. Furthermore, gases and volatile elements are also focused on. The subject of the following work is on Polonium, which becomes volatile at elevated temperatures.

The work presented serves primarily as a benchmark of codes which are used for calculating the nuclide inventory of radioactive material to be disposed in a final repository. These benchmarks help to estimate the reliability and uncertainties to be applied on the calculated nuclide inventory. However, a benchmark can only be reliably performed when experimental data can be used for comparison. This often requires detailed and complicated geometry, which 1) necessitates effort to implement into the code and 2) necessitates long running times to get enough statistics in a tiny region of an extended geometrical model. Therefore, most experimentally valuable data are usually compared to only one code. In this work, we use a simplified geometrical model for the benchmark which preserves the main feature of the extended model. At the same time, it averages the nuclide inventory in a much larger region compared to the samples used for the experimental examination. This allows reliable statistics to be acquired without stressing computer resources too hard. Finally, it will be shown that comparing the activities obtained in the toy model to the experimental data taken in a tiny lead sample of a complex geometry, can be justified.

**Results presented at SATIF-12**

In the SATIF-12 workshop [1], a comparison between experimentally determined activities of several isotopes in lead samples [2,3] from the Target 4 of the SINQ neutron spallation source [4] at the Paul Scherrer Institut (PSI) with results from the particle transport Monte Carlo calculations MCNPX2.7.0 [5] using different cross section models was presented. While several $\gamma$- and $\beta$-emitters could be predicted satisfactorily with the default model in MCNPX, the BERTINI-DRESNER-RAL model [6-9], discrepancies of up to 5 orders of magnitude to the experimental data occurred in case of the three polonium isotopes: Po-208, Po-209 and Po-210. In the case of Po-208 and Po-209, the reason was traced back to a deficit in the production by alpha capture on the stable lead isotopes. In the case of Po-210, the triton capture on Pb-208 forming Pb-210 decaying into Bi-210 and subsequently Po-210 was significantly suppressed. The large discrepancy in case of Po-210 only became obvious due to the measurement of the activities 10 years after the last beam on target. With a 138 day half-life, the directly produced Po-210 was mainly decayed, but the production due to the decay of Pb-210 with a half-life of 22 years was still significant. In both cases, the main reason for the significant discrepancy was the missing strength in the alpha and tritium production cross sections. Therefore, the corresponding alpha and tritium fluxes were smaller than obtained with INCL4.6/ABLA07 [10,11]. The activities calculated using INCL4.6/ABLA07 lead to very good agreement to the measured Po-activities.

The radioanalyses were performed at PSI [2,3] using five 1 mm slices of one of the 300 SINQ cannelloni (stainless steel tubes filled with lead). The cannellonis filled with Pb are stacked in 36 rows and separated by a few millimetres of deuterated water. The cannelloni used for the measurement was located in row 2. The deuterated water is used for removing the heat caused by the 575 MeV proton beam. To distribute the energy deposit ion, the beam shape is a double Gaussian with a standard deviation of several centimeters. This target, called Target4, was in operation from March 2000 until December 2001 and collected a charge of 10 Ah.

**The toy model**

A toy model was used in the following calculations to the benchmark using MCNPX2.7.0, GEANT4 and FLUKA. Its layout is shown in Figure 1. The model consists of 5 equal blocks of pure lead which are cladded with 1 mm of stainless steel. The blocks are surrounded by heavy water. The whole geometry, including the blocks, has a rectangular shape. The proton beam hits the ensemble from below in the center of the rectangle. The energy of the simulated proton beam is 575 MeV like it is in reality. However, it is just a very narrow pencil beam. The energy is less than the full 590 MeV from the accelerator since the protons passed through two meson production station and lose some energy before hitting the SINQ target. Since the nuclide inventory is averaged over each lead block and power density does not matter, the shape of the beam does not matter. The size of the rectangular cross section of the lead block is equal to the 12.4 cm x 12.4 cm rectangular shape of Target 4. Its real height of 60 cm is also used in the toy model. Each Pb block is 4.8 cm high. Further, the total amount of lead in Target 4 is equally distributed to the 5 blocks. The water in between the cannellonis is represented by the water between the blocks in the toy model. The 1 mm cladding resembles the cannelloni tube and is present in the real target so the heavy water is not in direct contact with lead, i.e. spallation products in the lead cannot react with the water. The inner sandwiched part of the toy model represents the entire SINQ Target 4. Both contain 43 % D2O and about 57 % Pb and steel.

**Figure 1: Geometry and materials of the toy model.**

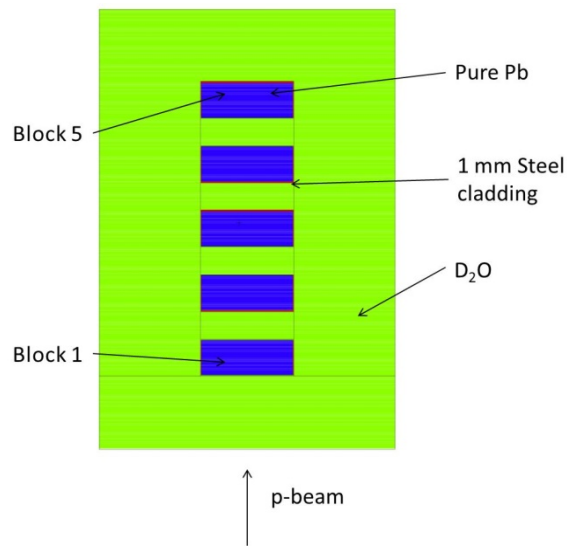

The advantage of the simplified SINQ model is that good statistics are much easier to aquire in a 12.4 cm x 12.4 cm x 4.8 cm lead block than in a 1 mm slice of 10 mm diameter – the size of one sample from the radioanalysis. The location of the samples experimentally analysed are from a region that corresponds to the center of Block 1. Therefore most of the results presented concentrates on this region.

The irradiation history corresponds to the original SINQ Target No. 4, i.e. 2 years of irradiation totaling 10 Ah and a shutdown period in between. Several cooling times were chosen. However, the results presented correspond to a cooling time of 10.6 years after end of beam. This is the time at which the lead samples were experimentally analysed and the activities of the Po isotopes determined.

**The particle transport codes**

The benchmark was performed with three different particle transport Monte Carlo codes: MCNPX2.7.0, GEANT4 and the newest, not yet published version of FLUKA, v2016.5. In the following sections, the three codes are described.

*MCNPX2.7.0*

MCNPX2.7.0 can transport all kind of particles: protons, neutrons, pions up to heavy nuclei. For their nuclear reactions, different physics models are used when no cross section tables are available or selected. Many physics models for the intermediate energy range (hundreds of MeV to a few GeV) make use of a microscopic picture of an intranuclear cascade (INC) followed by evaporation or fission. In the first step, the INC, the energy transferred by the primary particle is dissipated in the nucleus to several nucleons by nucleon-nucleon collisions. In the beginning of this phase, the so called pre-equilibrium, particles which receive a large energy transfer by direct interaction can leave the nucleus with large kinetic energy in forward direction (relative to the incoming particle) due to momentum conservation. After the remaining energy is distributed, the nucleus is in an excited state with often non-vanishing angular momentum. Both can be released either by evaporation or fission. Evaporation means that light particles, predominately neutrons and photons but also light ions, are isotropically emitted at low energy. Heavier nuclei underlay often surface vibrations, which lead finally to fission.

In the following evaluation, three physics models were used in MCNPX2.7.0:

- BERTINI-DRESNER-RAL [6-9]
- ISABEL[12]
- INCL4.6/ABLA07 [10-11]

For years, the INC model BERTINI [6,7] has been the default choice in MCNPX for neutrons and protons of energy less than 3.5 GeV. The BERTINI model, which uses a preequilibrium step by default, is coupled to the evaporation code of DRESNER [8]. The DRESNER code is based on Weisskopf´s statistical model. Fission, if possible, is handled by the RAL code [9]. Since BERTINI can handle only protons and neutrons, reactions containing light ions up to 1 GeV are handed over to ISABEL [12].

ISABEL can handle neutrons and protons as well as light ions, and therefore ISABEL can replace BERTINI completely. DRESNER and RAL still serve as evaporation and fission code in this case.

MCNPX2.7.0 contains an outdated version of INCL/ABLA, the INCL4.2/ABLAV3p, which has a couple of well-known shortcomings. Therefore, INCL was upgraded to INCL4.6 [10] and ABLA to ABLA07 [11]. However, this model was only implemented in MCNPX2.7.0 in a private version. The new

INCL4.6 model, which is based on INC according to its name, handles neutrons, protons, pions, deuterons, tritons, $^3$He and $\alpha$-particles, i.e. reaction with these particles is supported and these particles can also be emitted in all phases of the physics model. In addition, light fragments up to a mass number of 8 can be emitted during the intranuclear cascade phase. This is done by forming these clusters by coalescence in phase space, which is an important mechanism. ABLA07 itself is able to generate neutrons, protons, deuterons, tritons, $^3$He, $\alpha$-particles as well as intermediate mass fragments (IMF) by break-up, fission or evaporation. Although not relevant for the present work regarding the Po isotopes, the newest implementation of INCL4.6/ABLA07 in MCNPX2.7.0 also handles metastable isotopes according to the PHTLIB file in MCNPX. At the time of SATIF12, this was yet an issue.

In the following simulations, the above mentioned models are used for all particle interactions except for neutrons with energies less than 20 MeV. Here tabulated cross sections from the ENDF/B-VII [13] dataset are applied. In the cells selected for calculating the nuclide inventory, the neutron flux spectra up to 20 MeV are written to a text file. For all other interactions controlled by the physics models, a lot of information about each reaction (including the production rates for the residual nuclei) is usually written into a binary file called histp. This file can become several GB in size. Since the calculation of the nuclide inventory requires only the production rates, this information is stored in a text-file and the histp-file can be dismissed. This is possible using the rnucs-card, which was developed by F. Gallmeier and M. Wohlmuther [14]. With the help of a PERL script ("activation script" [15]) all information including the irradiation history and cooling times is passed to build-up and decay codes like CINDER1.05 [16] or FISPACT10 [17] using the cross section library EAF-2010 [17]. In the following evaluation, the results using FISPACT10/EAF-2010 are presented. However, a check with CINDER1.05, which comes with its own library, did not lead to significant difference in results.

For each MCNPX2.7.0 run using the three different physics model options, more than $10^9$ primary protons (= beam protons) hit the toy model. Details can be found in Table 1. The calculations were performed on a medium-performance cluster at PSI, which might be a factor 2 slower than high-performance clusters. The number of cores used was between 24 and 72. The core-hours per primary particle depend on the physics model chosen. The default model in MCNPX2.7.0, BERTINI-DRESNER-RAL, is the fastest one. The ISABEL model runs 6 % slower, INCL4.6/ABLA07 20 % slower. These numbers might differ a bit for another geometry used in the simulation.

**Table 1: Statistics and run time for MCNPX2.7.0 using the three different physics models.**

| MCNPX + physics model | No of primary protons | Cores *hours | Cores *hours/ nps |
|---|---:|---:|---:|
| BERTINI-DRESNER-RAL | 1.2 $10^9$ | 3658 | 3.05 $10^{-6}$ |
| ISABEL-DRESNER-RAL | 2.0 $10^9$ | 6912 | 3.46 $10^{-6}$ |
| INCL4.6/ABLA07 | 2.4 $10^9$ | 6912 | 2.88 $10^{-6}$ |

### GEANT4

Geant4 [18-20] is a toolkit for the simulation of the passage of particles through matter. It includes a complete set of functionality including tracking, geometry, physics models for creating an application to perform a Monte Carlo particle transport calculation. The latest public version of 10.02.p02 has been used for the application in this comparison. Geant4 results in this paper have been calculated with a reference physics list of QGSP_INCLXX_HP. INCL++, aka INCLXX is the C++ version of the Liège Intra-nuclear Cascade model. It generates final states for inelastic scattering of nucleons, pions and light nuclei (A<18) on nuclei in the physics list. De-excitation is handled by G4ExcitationHandler by default. The handler invokes evaporation, Generalized Evaporation Model, multi-fragmentation, Fermi breakup and the fission model of Geant4. Quark Gluon String Model is used for high-energy interactions (>15GeV).

Because of the energy of the primary proton, the high-energy model was never activated in the calculation. The ENDF data library based High Precision package is used for low energy (<20MeV) neutron transport. The RadioactiveDecay module is added to the physics list by physics constructor. This module decays unstable nuclide, including isomers, down to a stable isotope.

The calculation for this evaluation is performed in two steps: First, the energy spectra of tritons and alphas in the lead blocks of the toy model are calculated. Second, the production rates in the lead blocks are derived from the convolution of the particle spectra with the production cross sections. The production rates are then fed into Cinder07 [16] taking into account the irradiation profile.

Geant4 also provides other intra-nuclear cascade models. Results from them are provided in [21].

### FLUKA2016.5

FLUKA [22] version 2016.5 [23-24] has been used for this comparison. The nuclear models of relevance for elementary hadron nuclear interactions and for alpha and triton nuclear interactions are, respectively, the PEANUT [25] hadron-nucleus and the BME [26] nucleus-nucleus reaction models.

PEANUT is a Generalized IntraNuclear Cascade model able to simulate hadron-nucleus, photo-nuclear, electro-nuclear, and neutrino-nucleus interactions from threshold, from 20 MeV on for neutrons, until several TeV laboratory energy. It includes a preequilibrium step and a coalescence algorithm, which is obviously of great relevance for the results presented in this paper. Both models share the standard FLUKA evaporation/fragmentation/fission model.

The most recent developments for both models can be found in [23-24]. BME in particular has been extensively improved and now includes the interfacing with the PEANUT preequilibrium for all reaction channels not included in the original BME approach.

**Comparison of the alpha spectra and cross sections**

Since alphas play a major role in the production of Po-208 and Po-209, the alpha flux in block 1 was compared for the different codes/models as a function of energy up to about 150 MeV. The results are seen in Figure 2. For the comparison, the same bin sizes and a comparable statistics were used. In addition, the fluxes were normalized to the energy bin size.

**Figure 2 : Alpha flux spectrum for the different models in block 1 of the toy model.**

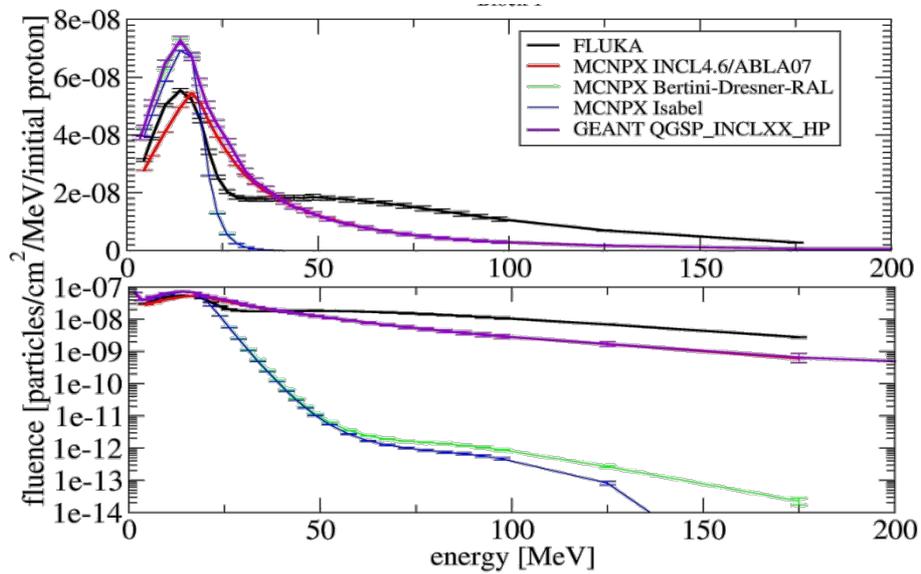

The upper part of Figure 2 shows the results in linear scale to make differences in the dominant peak more visible. MCNPX using BERTINI or ISABEL alone leads to almost identical results. This is not surprising since in case of BERTINI, the ISABEL code also handles the alpha particles. GEANT using INCL leads to an almost identical result in the peak compared to BERTINI/ISABEL, whereas the high energy tail of the alpha flux cannot be distinguished from the result using MCNPX INCL4.6/ABLA07. FLUKA is the only code whose alpha spectrum has two peaks. One peak is at the same location as the other codes/models, and the second peak is very broad and at higher energies where the other results using INCL have just a tail. The tail can be compared well in the lower part of Figure 2 in a logarithmic scale. Due to the second peak, FLUKA has the largest strength in the high energy tail. This tail is almost completely missing in the BERTINI/ISABEL code, i.e. the high energetic alpha emission in forward direction is suppressed. In the modern codes like INCL4.6 and PEANUT in FLUKA, high energy light ions are emitted via the mechanism of coalescence. This allows forming light ions out of their ingredients, i.e. neutrons and protons that are "glued" together.

**Figure 3 : Comparison of the cross sections obtained with different codes and physics models for the production of Po-208 by alpha captures from Pb-206.**

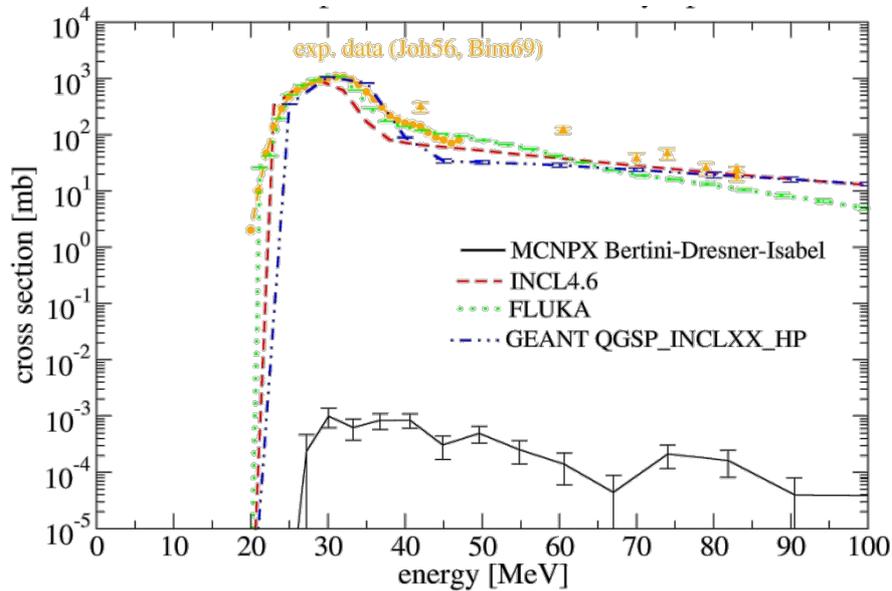

Since the production of the Po-isotopes depends on both the presence of the alphas, i.e. the alpha flux spectrum, and the production cross section, the cross section for the production of Po-208 by alpha capture on the stable lead isotope Pb-206 is shown in Figure 3. The cross section for INCL4.6 is compared to the cross sections extracted from MCNPX BERTINI/ISABEL, FLUKA and GEANT using INCL. Except for BERTINI-DRESNER, the other cross section models lead to very similar results, which also agree with the experimental database [27-28]. In case of BERTINI-DRESNER about 6 orders of magnitude is missing. The situation for the production of Po-209 and Po-210 by alphas from stable lead isotopes is very similar to the presented case.

**Comparison of the triton spectra and cross sections**

As for alphas, the triton flux spectrum in block on of the toy model is shown for the different codes and physics models in Figure 4. The upper plot of the figure is in linear energy scale and the lower is in logarithmic energy scale. The triton flux obtained with MCNPX BERTINI/ISABEL contains almost only low energy particles. The high energy tail is missing. Compared to the results from the other models, even the low energy tritons are significantly suppressed. The results from MCNPX and GEANT both with INCL are very similar in shape and magnitude in the low as well as the high energy part. From around 50 MeV on, the spectra from MCNPX INCL, GEANT INCL and FLUKA are almost identical.

**Figure 4 : Triton flux spectrum for the different models in block 1 of the toy model.**

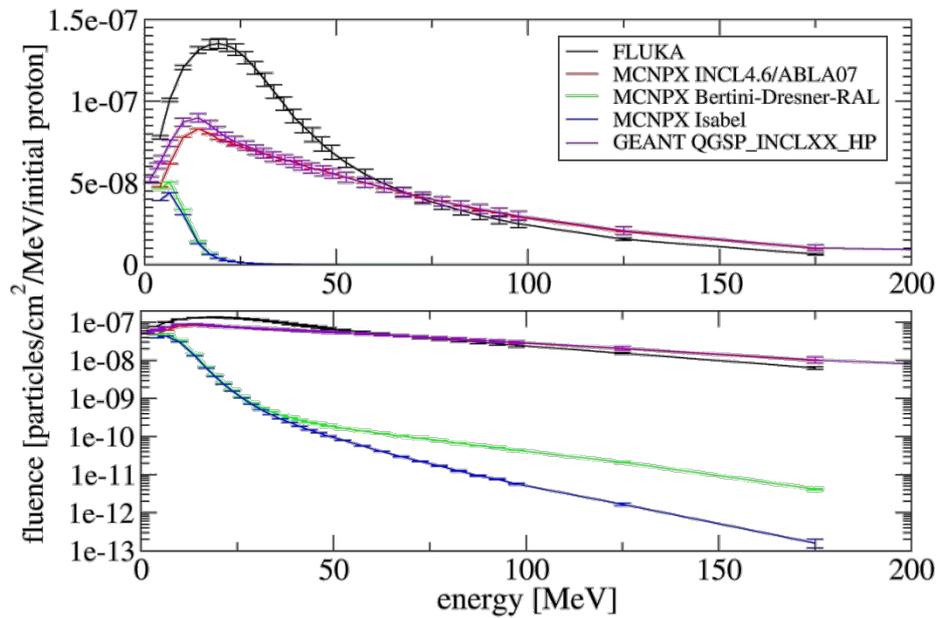

**Figure 5 : Comparison of the cross sections obtained with different codes and physics models for the production of Pb-210 by triton capture from Pb-208.**

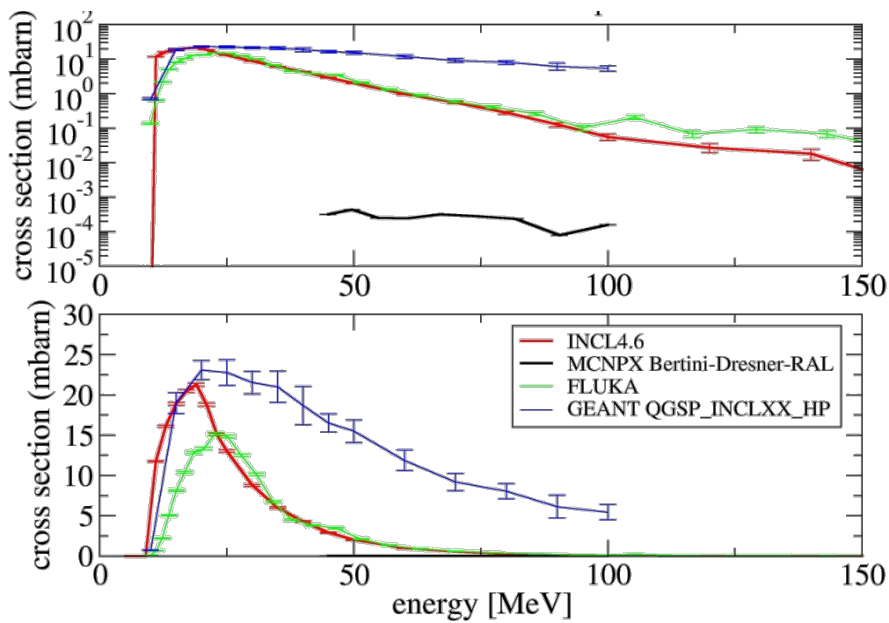

The triton flux is relevant for the production of Po-210 via the triton capture on the stable lead isotope Pb-208, which leads to Pb-210. This lead isotope decays with a half-life of 22 year into Bi-210 and finally into Po-210. Therefore, the production cross section Pb-208(t,X)Pb-210 is compared in Figure 5 for the different codes and physics models. The upper plot contains the results in linear energy scale and the lower plot is in logarithmic energy scale. The PEANUT model and INCL4.6 agree very well to each other. The INCL model in GEANT4 agrees well with MCNPX/INCL4.6 up to the peak.

For higher energy tritons, the cross section extracted from GEANT4/INCL4.6 is about a factor 2 larger than MCNXP/INCL4.6 and FLUKA. The production cross section produced by MCNPX/BERTINI is about 4 orders of magnitude lower.

**Comparison of the Po-activities**

In the 5 blocks of the toy model, the specific activities of the Po isotopes have compared for the different codes /models after 10.6 years of cooling time after irradiation. The results are shown for Po-208 in Figure 6. As expected, the Po activity decreases with the block number, i.e. with the energy of the proton beam. The results obtained for FLUKA, MCNPX-INCL4.6/ABLA07 and GEANT4-INCL agree very well, particularly for the blocks 1-4. The deviations are larger in the last block. A likely reason is the poorer statistics at the upper end of the target. The results from MCNPX-BERTINI/ISABEL are suffering from statistics as the activities are 2 to 3 orders of magnitude smaller.

**Figure 6. Comparison of the Po-208 activities in the 5 blocks of the toy model using different codes and physics models.**

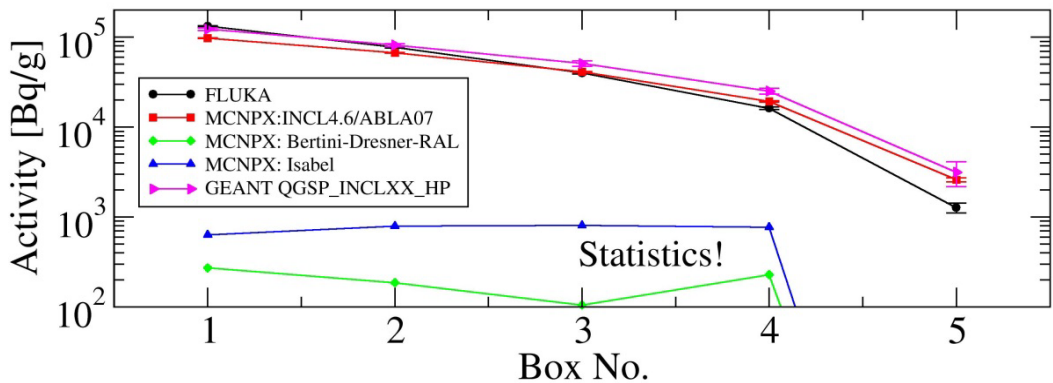

The comparison of the results for the Po-209 activities in Figure 7 leads to the same conclusions made for the Po-208 activities.

**Figure 7 : Comparison of the Po-209 activities in the 5 blocks of the toy model using different codes and physics models.**

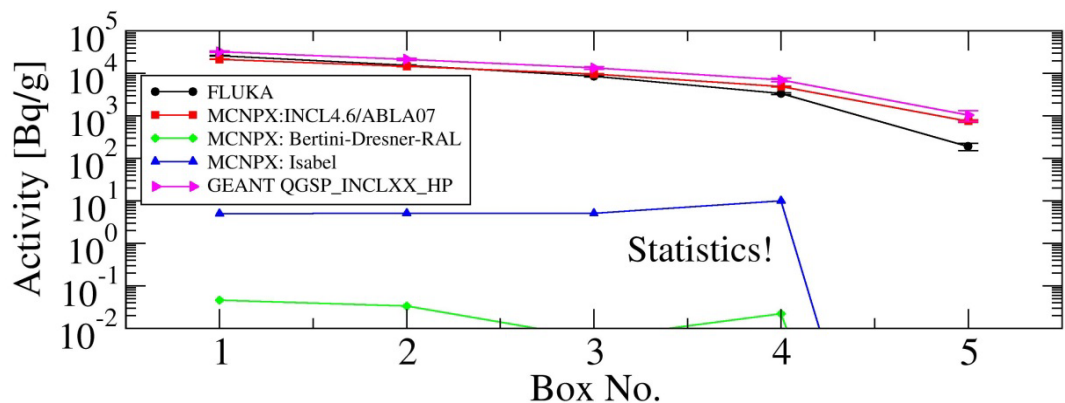

**Figure 8 : Comparison of the Po-210 activities in the 5 blocks of the toy model using different codes and physics models.**

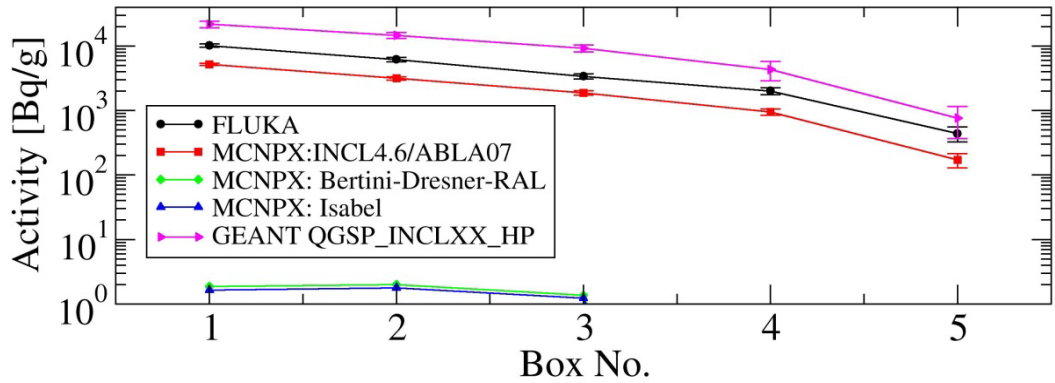

In Figure 8, the results for the Po-210 specific activities from the different codes/models are compared for the five blocks. With respect to the activities for the two Po isotopes compared in Figure 6 and 7, the spread of the results are largest for the Po-210 activities. Neglecting the much too small result of MCNPX-BERTINI/ISABEL, the FLUKA result seems to represent a kind of average for GEANT4-INCL and MCNPX-INCL4.6/ABLA07. Contrary to the activities of the other two Po isotopes, the deviation of the results from the three different codes seems to be almost constant (about a factor 2). However, the production channel via triton capture is the more difficult and exotic reaction.

**Figure 9 : Ratio of the Po activities from block 1 of the toy model to the result from MCNPX-INCL4.6/ABLA07.**

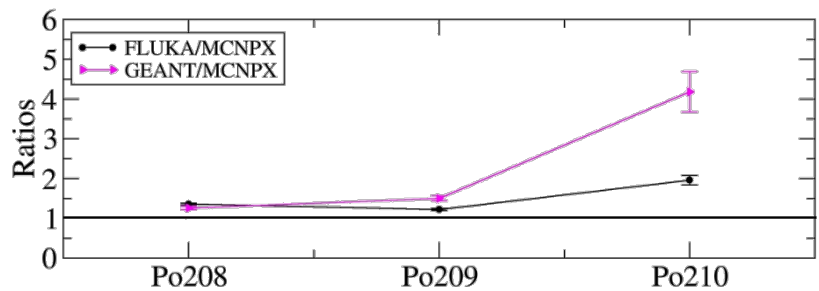

For a better comparison, the ratios of the activities for the three Po isotopes calculated with FLUKA and GEANT4-INCL to the results from MCNXP-INCL4.6/ABLA07 are shown for block 1 and 10.6 years of cooling time. The deviation between FLUKA and MCNPX model is between 1.2 and 2, whereas the difference between GEANT4 and MCNPX model is between 1.2 and 4. GEANT4 leads to the more conservative results.

**Comparison with experimental data**

The aim of the benchmark is not only to compare the results obtained in different particle transport Monte codes and physics models, but also to compare them experimental data. However, the experimental data were taken in a complex geometry, whereas the toy model is a very simplified version of it. Since the alpha and triton spectra play a decisive role (production cross sections do not depend on the geometrical model) these spectra are compared for both models. In Figure 10, the alpha and triton flux spectra in Rod 3 of the complex SINQ model are compared with the corresponding flux spectra in block 1 of the toy model. The samples for the experimental investigations were taken from Rod 3. Rod 3 corresponds to the center of block 1 in the toy model. The spectra in Figure 10 are scaled to each other as the lead volume is quite different. Besides this, the alpha spectra (Figure 10 left) as well as the triton spectra (Figure 10 right) are almost identical in both geometrical regions of the models.

**Figure 10 : Comparison of the alpha and triton spectra in block 1 of the toy model and Rod 3 of the detailed SINQ model.**

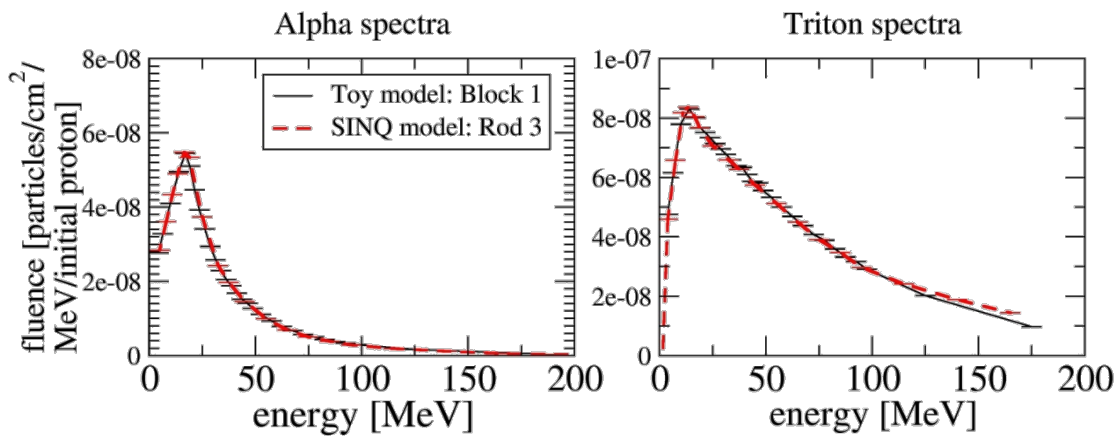

The good agreement of the alpha and triton spectra justifies comparing the calculated Po activities in the toy model with the measured Po activities.

**Figure 11 : Comparison of the calculated activities in the toy model to the experimental data.**

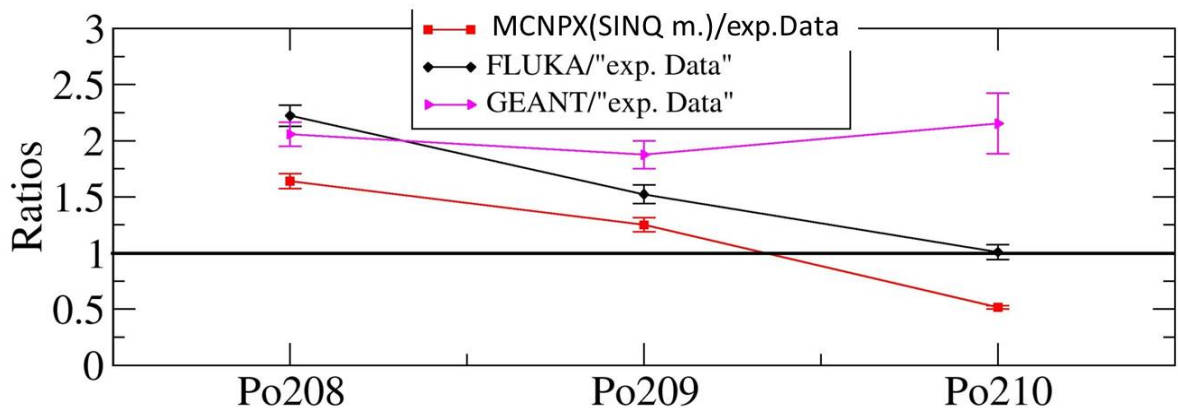

For ease of comparison, the ratio from the evaluation presented at SATIF12 is shown in Figure 11 again (red line). In this ratio, the specific activities calculated with MCNPX2.7.0-INCL4.6/ABLA07 in the most realistic SINQ model at the position of the samples is divided by the experimental result in these samples. Then the ratios shown in Figure 9, i.e. FLUKA as well as GEANT divided by the MCNPX toy model results, is multiplied by the values of the red line in Figure 11, i.e. the ratio of the MCNPX SINQ model to the experimental data.  These two lines are also shown in Figure 11. They represent the comparison of the results from FLUKA and GEANT using the toy model to the experimental data. Comparing the ratio with FLUKA and GEANT to the one from MCNPX, the results with FLUKA and GEANT are larger for all three Po isotopes, i.e. more conservative than the MCNPX results. The best agreement for Po-210 is achieved with FLUKA. The deviation from the experimental results is maximum a factor of 2. GEANT shows an almost constant factor of 2 deviations for roughly all three Po isotopes.

**Summary and Conclusion**

The focus of the present evaluation is to compare different codes and physics models in the primary reaction of 575 MeV protons on bulk lead. The codes in this benchmark are MCNPX2.7.0, GEANT4 and FLUKA2016.5, the newest version not yet available for the public. The benchmark compares the specific activities of Po-208, Po-209 and Po-210 calculated from the simulation results. The dominant reaction mechanism are alpha and triton capture on stable lead isotopes. The triton capture is important for Po-210 after a cooling time of several years. Then the directly produced Po-210 with a half-life of 138 days is already decayed, whereas Pb-210, which was produced by triton capture on Pb-208, feeds Po-210 via the decay of Bi-210 and has an effective half-life of 22 years. Therefore, the alpha and triton flux spectra, as well as the alpha and triton production cross sections, were compared as well.

Good agreement between the codes using modern physics models like PEANUT in FLUKA and INCL4.6/ABLA07 in MCNPX2.7.0 and GEANT4 was achieved. However, physics models like BERTINI-DRESNER-RAL and ISABEL lead to deviations of up to 4 orders of magnitude. This could be traced back to the alpha and triton production cross sections, which are negligible in magnitude compared to PEANUT and INCL4.6, as well as to the alpha and triton spectra. In the spectra, the long tail of high-energy particles is missing in both spectra. The reason for this discrepancy is the missing coalescence mechanism in the BERTINI and ISABEL models.

Finally, the results obtained in a simplified toy model of the SINQ facility at PSI were compared to the experimentally determined activities of Po-208, Po-209 and Po-210 measured in samples taken in SINQ Target 4. Since the alpha and triton spectra of block 1 of the toy model and of the location of the samples are almost identically, when scaled to each other, the comparison is anticipated as justified. The three codes using modern physics models deviate at maximum a factor of 2 from the experimental data. This good agreement is really astonishing as the production mechanisms of all three Po isotopes are non-trivial.

**Acknowledgement**

We thank Dorothea Schumann and Tobias Lorenz for their very helpful collaboration and for providing the experimental results for this comparison.